\documentclass[twocolumn,aps,prl,final,superscriptaddress]{revtex4-1}
\usepackage[latin9]{inputenc}
\usepackage[active]{srcltx}
\usepackage{color}
\usepackage{amsmath}
\usepackage{graphicx}
\usepackage[unicode=true,pdfusetitle,
 bookmarks=true,bookmarksnumbered=true,bookmarksopen=true,bookmarksopenlevel=1,
 breaklinks=true,pdfborder={0 0 1},backref=false,colorlinks=true]
 {hyperref}
\hypersetup{
 linkcolor=blue,urlcolor=blue,citecolor=blue,pdfstartview={FitH},hyperfootnotes=false}

\makeatletter

\newcommand*\LyXThinSpace{\,\hspace{0pt}}

\usepackage{dcolumn}
\usepackage{bm}

\usepackage{times}

\makeatother

\begin{document}
\title{Linear\textit{ }versus nonlinear electro-optic effects in materials}
\author{Zhijun Jiang}
\affiliation{Key Laboratory of Computational Physical Sciences (Ministry of Education),
State Key Laboratory of Surface Physics, and Department of Physics,
Fudan University, Shanghai 200433, China}
\affiliation{Physics Department and Institute for Nanoscience and Engineering,
University of Arkansas, Fayetteville, Arkansas 72701, USA }
\affiliation{School of Physics and Optoelectronic Engineering, Ludong University,
Yantai 264025, China}
\author{Charles Paillard}
\affiliation{Physics Department and Institute for Nanoscience and Engineering,
University of Arkansas, Fayetteville, Arkansas 72701, USA }
\affiliation{Laboratoire SPMS, CentraleSupélec/CNRS UMR 8580, Université Paris-Saclay,
8-10 rue Joliot Curie, 91190 Gif-sur-Yvette, France}
\author{Hongjun Xiang}
\email{hxiang@fudan.edu.cn}

\affiliation{Key Laboratory of Computational Physical Sciences (Ministry of Education),
State Key Laboratory of Surface Physics, and Department of Physics,
Fudan University, Shanghai 200433, China}
\affiliation{Collaborative Innovation Center of Advanced Microstructures, Nanjing
210093, China}
\author{L. Bellaiche}
\email{laurent@uark.edu}

\affiliation{Physics Department and Institute for Nanoscience and Engineering,
University of Arkansas, Fayetteville, Arkansas 72701, USA }
\begin{abstract}
Two schemes are proposed to compute the \textit{nonlinear} electro-optic
(EO) tensor for the first time. In the first scheme, we compute the
linear EO tensor of the structure under a finite electric field, while
we compute the refractive index of the structure under a finite electric
field in the second scheme. Such schemes are applied to Pb(Zr,Ti)O$_{3}$
and BaTiO$_{3}$ ferroelectric oxides. It is found to reproduce a
recently observed feature, namely why Pb(Zr$_{0.52}$Ti$_{0.48}$)O$_{3}$
adopts a mostly linear EO response while BaTiO$_{3}$ exhibits a strongly
nonlinear conversion between electric and optical properties. Furthermore,
the atomistic insight provided by the proposed \textit{ab-initio}
scheme reveals the origin of such qualitatively different responses,
in terms of the field-induced behavior of the frequencies of some
phonon modes and of some force constants. 
\end{abstract}
\maketitle
Most materials exhibit a change in their refractive index when under
applied static or low-frequency electric fields, a phenomenon known
as the electro-optic (EO) effect \cite{Weber2002,Lines1997} and which
is promising for some technologies \cite{Zgonik1994,Veithen2004,Chen2014,Charles2019}.
In particular, having large \textit{nonlinear} electro-optic coefficients
would open the door for the design of novel devices \cite{Wemple1972,Saleh1991,Yariv2007,Boyd2008,DiDomenico1969,Wemple1969,Kuzyk1990,Hisakado2005,Jin2012,Shen2014}.
For instance, it is important for EO modulation \cite{Kuzyk1990},
high-speed optical shutters \cite{Hisakado2005}, electro-optical
detection \cite{Jin2012}, and electro-optical switching \cite{Shen2014}.
Understanding at an atomistic level linear \textit{versus} nonlinear
EO effects should also be of large fundamental interest. For instance,
it should resolve the current mystery of why a recent experiment \cite{Chen2014}
observed, in the THz regime, a linear electro-optic coefficient in
Pb(Zr,Ti)O$_{3}$ while BaTiO$_{3}$ films rather exhibit significant
nonlinear (second-order) electro-optic coefficients.

In view of such facts, having a first-principles-based technique allowing
the computation of nonlinear conversion between electric and optical
quantities but also providing a deep atomistic insight of such conversion
is highly desired. However, such technique and \textit{ab-initio}
capabilities do not presently exist.

The aims of this paper are to demonstrate that such technique (1)
can, in fact, be easily developed and applied to any material; (2)
reproduces the experimental finding about the different nature (i.e.,
linear versus nonlinear) of the EO response of Pb(Zr,Ti)O$_{3}$ \textit{versus}
BaTiO$_{3}$; and (3) explains such difference, via the field-induced
behavior of some specific phonon frequencies and of the force constants
of some chemical bonds.

Here, we employed the ABINIT package \cite{Gonze2002} with the local
density approximation (LDA) to the density functional theory (DFT)
and norm-conserving pseudopotentials \cite{Hamann2013}, chosen in
part to facilitate the computation of electro-optic coefficients \cite{Veithen2004,Veithen2005}.
The alchemical mixing approximation implemented in the ABINIT package
\cite{Scharoch2014}, which is the pseudopotentials specific implementation
of the virtual crystal approximation, is also adopted to investigate
the Pb(Zr$_{0.52}$Ti$_{0.48}$)O$_{3}$ (PZT) solid solutions. We
use a 8 $\times$ 8 $\times$ 8 grid of special \textbf{k} points
and a plane-wave kinetic energy cutoff of 50 hartree. The effects
of $dc$ electric fields applied along the {[}111{]} direction on
structural properties of the rhombohedral $R3m$ phase of both BaTiO$_{3}$
(BTO) and PZT are simulated by taking advantage of the method developed
in Refs. \cite{Nunes1994,Nunes2001,Souza2002,Zwanziger2012} (note
that the $R3m$ phase is the well-known ground state of BTO, and that
we chose to study a Ti composition of 48\% in PZT in order to have
a stable rhombohedral ferroelectric state as well). Technically, for
each considered magnitude of the $dc$ electric field, both the lattice
parameters and the atomic positions were fully relaxed until the force
acting on each atom is smaller than 5 $\times$ 10$^{-5}$ hartree/bohr.
The resulting field-induced structures are then used as input for
the \textit{ab-initio} method of Refs. \cite{Veithen2004,Veithen2005,Jiang2019},
that is based on the linear response of the optical dielectric tensor
induced by a static (or low frequency) electric field $E_{k}$ and
that has been proven to accurately compute EO coefficients in ferroelectric
oxides (note that no electric field is incorporated when employing
this latter method on the field-induced structures). Such coefficients
obey the following equation 
\begin{equation}
\Delta(\varepsilon^{-1})_{ij}=\sum_{k=1}^{3}{\mathcal{R}}_{ijk}^{\mathrm{\eta}}E_{k},\label{eq:EO tensor}
\end{equation}
where $(\varepsilon^{-1})_{ij}$ is the inverse of the electronic
dielectric tensor that depends on the electric field. It is important
to realize that, in our case, ${\mathcal{R}}_{ijk}^{\mathrm{\eta}}$
is a clamped (strain-free) EO tensor that can practically depend on
$E_{k}$ since we used the crystal structure spanned by this electric
field for its calculation. In particular, plotting ${\mathcal{R}}_{ijk}^{\mathrm{\eta}}$
\textit{versus} $E_{k}$ will naturally determine if the materials
under investigation only adopt a linear EO effect (in that case, ${\mathcal{R}}_{ijk}^{\mathrm{\eta}}$
will be independent of $E_{k}$) or rather a nonlinear conversion
between electric and optical quantities (which will make ${\mathcal{R}}_{ijk}^{\mathrm{\eta}}$
dependent on $E_{k}$).

As detailed in Refs. \cite{Veithen2004,Veithen2005}, ${\mathcal{R}}_{ijk}^{\mathrm{\eta}}$
can be expressed as the sum of two contributions: a bare electronic
part, ${\mathcal{R}}_{ijk}^{\mathrm{el}}$, which is proportional
to the nonlinear optical dielectric susceptibility $\chi_{ijk}^{(2)}$,
and an ionic part, ${\mathcal{R}}_{ijk}^{\mathrm{ion}}$, which is
caused by the relaxation of the atomic positions due to the applied
electric field and which depends on the first-order change of the
linear dielectric susceptibility. The origin of the ionic contribution
is related to the Raman susceptibility $\alpha_{ij}^{m}$ of mode
$m$, the transverse optic mode polarity $p_{k}^{m}$ and phonon mode
frequencies $\omega_{m}$. The clamped (strain-free) EO tensor is
thus given by: 
\begin{equation}
{\mathcal{R}}_{ijk}^{\mathrm{\eta}}={\mathcal{R}}_{ijk}^{\mathrm{el}}+{\mathcal{R}}_{ijk}^{\mathrm{ion}}=\frac{-8\pi}{n_{i}^{2}n_{j}^{2}}\chi_{ijk}^{(2)}-\frac{4\pi}{n_{i}^{2}n_{j}^{2}\sqrt{\Omega_{0}}}\sum_{m}\frac{\alpha_{ij}^{m}p_{k}^{m}}{\omega_{m}^{2}},\label{eq:EO tensor clamped}
\end{equation}
where $n_{i}$ and $n_{j}$ are the principal refractive indices,
and $\Omega_{0}$ is the unit cell volume. As taken advantage in previous
works \cite{Veithen2004,Charles2019,Veithen2005,Jiang2019}, Eq. (\ref{eq:EO tensor clamped})
can be used to provide a deep insight into EO coefficients. Examples
include the determination of the modes $m$ mostly responsible for
the value of these coefficients as well as their enhancement via the
softening of these modes (i.e., $\omega_{m}$ approaching a zero value).
Note also that there is an \textit{unclamped} (stress-free that adds
a contribution involving elasto-optic effects and piezoelectricity
to the clamped one) EO tensor, that is given by ${\mathcal{R}}_{ijk}^{\mathrm{\sigma}}={\mathcal{R}}_{ijk}^{\mathrm{\eta}}+\sum_{\alpha,\beta=1}^{3}p_{ij\alpha\beta}d_{k\alpha\beta}$
where $p_{ij\alpha\beta}$ are elasto-optic coefficients and $d_{k\alpha\beta}$
are piezoelectric strain coefficients \cite{Veithen2004,Charles2019,Veithen2005,Jiang2019},
but that we numerically found (see Fig. S1 of the Supplemental Material
(SM) \cite{SM}) that ${\mathcal{R}}_{ijk}^{\mathrm{\sigma}}$ and
${\mathcal{R}}_{ijk}^{\mathrm{\eta}}$ behave in a similar qualitative
and even quantitative way with the applied electric field, in both
PZT and BTO. Consequently, we focus here on ${\mathcal{R}}_{ijk}^{\mathrm{\eta}}$.
Note also that, as detailed in the SM \cite{SM}, we also computed
the EO tensor associated with the aforementioned field-induced structures
by using another (more brute force) method -- since we are not aware
that nonlinear electro-optic effects have ever been investigated using
first-principles-based calculations. This latter method and the one
explained above provide very similar results, which therefore attests
of the validity of the approach adopted in this manuscript.

\begin{figure}
\includegraphics[width=8cm]{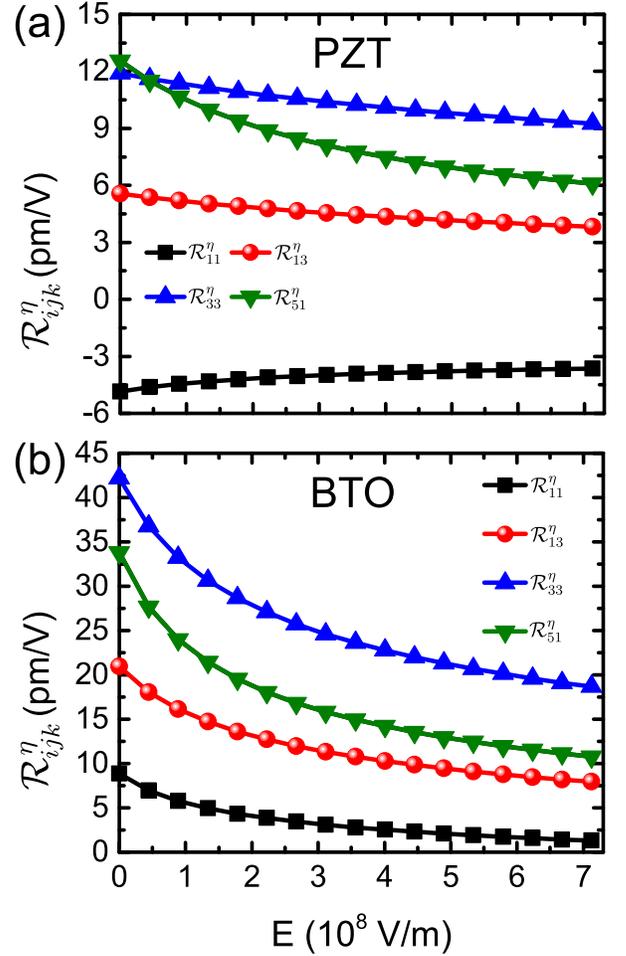}

\caption{Clamped EO coefficients as a function of electric field applied along
the {[}111{]} direction in (a) PZT and (b) BTO, respectively. \label{fig:clamped EO tensor}}
\end{figure}

Let us now choose the Cartesian axes such as the $z$-axis is along
the {[}111{]} polarization pseudo-cubic direction and the $y$-axis
is perpendicular to the mirror plane of the $R3m$ structure, for
both PZT and BTO. With this choice of axes and when adopting the Voigt
notation, the ${\mathcal{R}}_{ijk}^{\mathrm{\eta}}$ EO tensor has
four independent elements: ${\mathcal{R}}_{11}^{\mathrm{\eta}}$,
${\mathcal{R}}_{13}^{\mathrm{\eta}}$, ${\mathcal{R}}_{33}^{\mathrm{\eta}}$
and ${\mathcal{R}}_{51}^{\mathrm{\eta}}$. Figures~\ref{fig:clamped EO tensor}(a)
and \ref{fig:clamped EO tensor}(b) show all these components as a
function of the electric field applied along the {[}111{]} direction
in PZT and BTO, respectively, as calculated from Eq.~(\ref{eq:EO tensor clamped}).
Regarding PZT, Fig.~\ref{fig:clamped EO tensor}(a) indicates that
the clamped EO coefficients are predicted to be, at zero electric
field and by order of increasing strength, ${\mathcal{R}}_{11}^{\eta}$
$=$ $-$4.9 pm/V, ${\mathcal{R}}_{13}^{\eta}$ $=$ 5.6 pm/V, ${\mathcal{R}}_{33}^{\eta}$
$=$ 11.9 pm/V and ${\mathcal{R}}_{51}^{\eta}$ $=$ 12.6 pm/V (note
that we are not aware of any measurement or calculation of these coefficients
in rhombohedral PZT). When the electric field is turned on and increases
in PZT, ${\mathcal{R}}_{33}^{\eta}$ quickly becomes the largest element
and basically only very slightly \textit{linearly} decreases, therefore
indicating the occurrence of a predominant linear EO coefficient accompanied
by a weak second-order EO response. In fact, one can nicely fit ${\mathcal{R}}_{33}^{\eta}$
by $r_{33}^{\eta}+s_{333}^{\eta}E_{3}$, which provides a linear (Pockels
effect) EO coefficient of $r_{33}^{\eta}$ $=$ 11.6 pm/V and a quadratic
(Kerr effect) EO parameter of $s_{333}^{\eta}$ $=$ $-$3.6$\times$
10$^{-21}$ m$^{2}$/V$^{2}$. Our predicted $r_{33}^{\eta}$ at 0
K is of the same order of magnitude and rather consistent with the
measurement of Ref. \cite{Chen2014} giving a value of 67.3 pm/V at
room temperature for Pb(Zr$_{0.52}$Ti$_{0.48}$)O$_{3}$ at an applied
$ac$ frequency of 1 THz, when realizing that temperature decreases
the soft-mode frequencies and therefore enhances EO coefficient --
as clearly indicated by Eq. ~(\ref{eq:EO tensor clamped}).

For BTO at zero field, all the clamped EO coefficients are larger
in magnitude than those of PZT, as shown in Fig.~\ref{fig:clamped EO tensor}(b)
that reports a value of ${\mathcal{R}}_{11}^{\eta}$ $=$ 8.9 pm/V,
${\mathcal{R}}_{13}^{\eta}$ $=$ 21.0 pm/V, ${\mathcal{R}}_{33}^{\eta}$
$=$ 42.2 pm/V and ${\mathcal{R}}_{51}^{\eta}$ $=$ 33.8 pm/V (note
that these values are consistent with the previously reported ones
of Ref. \cite{Veithen2005} but using the experimental lattice constants
of BTO). Moreover and in sharp contrast with PZT, Fig.~\ref{fig:clamped EO tensor}(b)
also reveals that all the elements of ${\mathcal{R}}_{ijk}^{\mathrm{\eta}}$
in BTO strongly depends on the magnitude of the electric field. Such
numerical finding is fully in-line with a recent experiment \cite{Chen2014}
observing a predominantly linear EO response in PZT films \textit{versus}
a nonlinear electro-optic response of BTO thin films, in the THz frequency
range. In fact, we numerically find that our computed ${\mathcal{R}}_{33}^{\eta}$
of BTO of Fig.~\ref{fig:clamped EO tensor}(b) can be very well fitted
by $r_{33}^{\eta}+s_{333}^{\eta}E_{3}+c_{3333}^{\eta}E_{3}^{2}$,
with $r_{33}^{\eta}$ $=$ 39.6 pm/V, $s_{333}^{\eta}$ $=$ $-$6.4$\times$
10$^{-20}$ m$^{2}$/V$^{2}$ and $c_{3333}^{\eta}$ $=$ 5.1 $\times$
10$^{-29}$ m$^{3}$/V$^{3}$. Note, however, that the magnitude of
the predicted second-order EO coefficient of 6.4$\times$ 10$^{-20}$
m$^{2}$/V$^{2}$ is about 200 times smaller in magnitude than that
measured in Ref. \cite{Chen2014} for an $ac$ frequency of 1 THz
(note that the other method described in the SM \cite{SM} does not
rely on linear response and provides similar result for the EO coefficients).
Possible reasons behind such discrepancy is that we study here the
$R3m$ phase at 0 K while experiments are conducted on the tetragonal
phase of BTO at room temperature, that is very close (namely by about
20 K \cite{Lines1997,Kwei1993,Walizer2006}) to the tetragonal-to-orthorhombic
phase transition where a large enhancement of the EO responses is
expected \cite{Veithen2005-1} due to the softening of some phonon
frequencies -- as evidenced by Eq. ~(\ref{eq:EO tensor clamped}).
Other possible reasons may be that \textit{ab-initio} electric fields
can typically provide an overestimation by one or two orders of magnitude
with respect to experimental ones \cite{Xu2017,Jiang2018,Lu2019,Chen2019},
or that the experiment in Ref. \cite{Chen2014} is conducted on strained
and unpoled samples while we study bulk polar materials. The latter
such hypothesis is even more reasonable when assuming the formula
indicated in Refs. \cite{Wemple1972,DiDomenico1969,Veithen2005-1},
that are: ${\mathcal{R}}_{13}^{\mathrm{0}}=\frac{2}{3}\left(g_{11}+2g_{12}-g_{44}\right)P_{s}\chi_{33}$,
${\mathcal{R}}_{33}^{\mathrm{0}}=\frac{2}{3}\left(g_{11}+2g_{12}+2g_{44}\right)P_{s}\chi_{33}$,
and ${\mathcal{R}}_{51}^{\mathrm{0}}=\frac{2}{3}\left(g_{11}-g_{12}+\frac{1}{2}g_{44}\right)P_{s}\chi_{11}$,
where ${\mathcal{R}}_{ij}^{\mathrm{0}}$ are linear EO coefficients,
$P_{s}$ is the spontaneous polarization, $\chi_{33}$ and $\chi_{11}$
are the dielectric constants along the $c$- and $a$-axes, respectively,
and $g_{ij}$ are specific quadratic EO coefficients -- all under
infinitesimally small electric fields. As a matter of fact, plugging
our numerical values for ${\mathcal{R}}_{ij}^{\mathrm{0}}$, $P_{s}$,
$\chi_{33}$ and $\chi_{11}$ into these formula gives for the $R3m$
phase of BTO bulk: $g_{11}$ $=$ 273, $g_{12}$ $=$ $-$2 and $g_{44}$
$=$ 68 in units of 10$^{-2}$ m$^{4}$/C$^{2}$, which is precisely
the order of magnitude reported in Ref. \cite{Wemple1972} for the
$g_{ij}$ coefficients of BTO bulk at room temperature (i.e., in the
$P4mm$ phase).

\begin{figure}
\includegraphics[width=8cm]{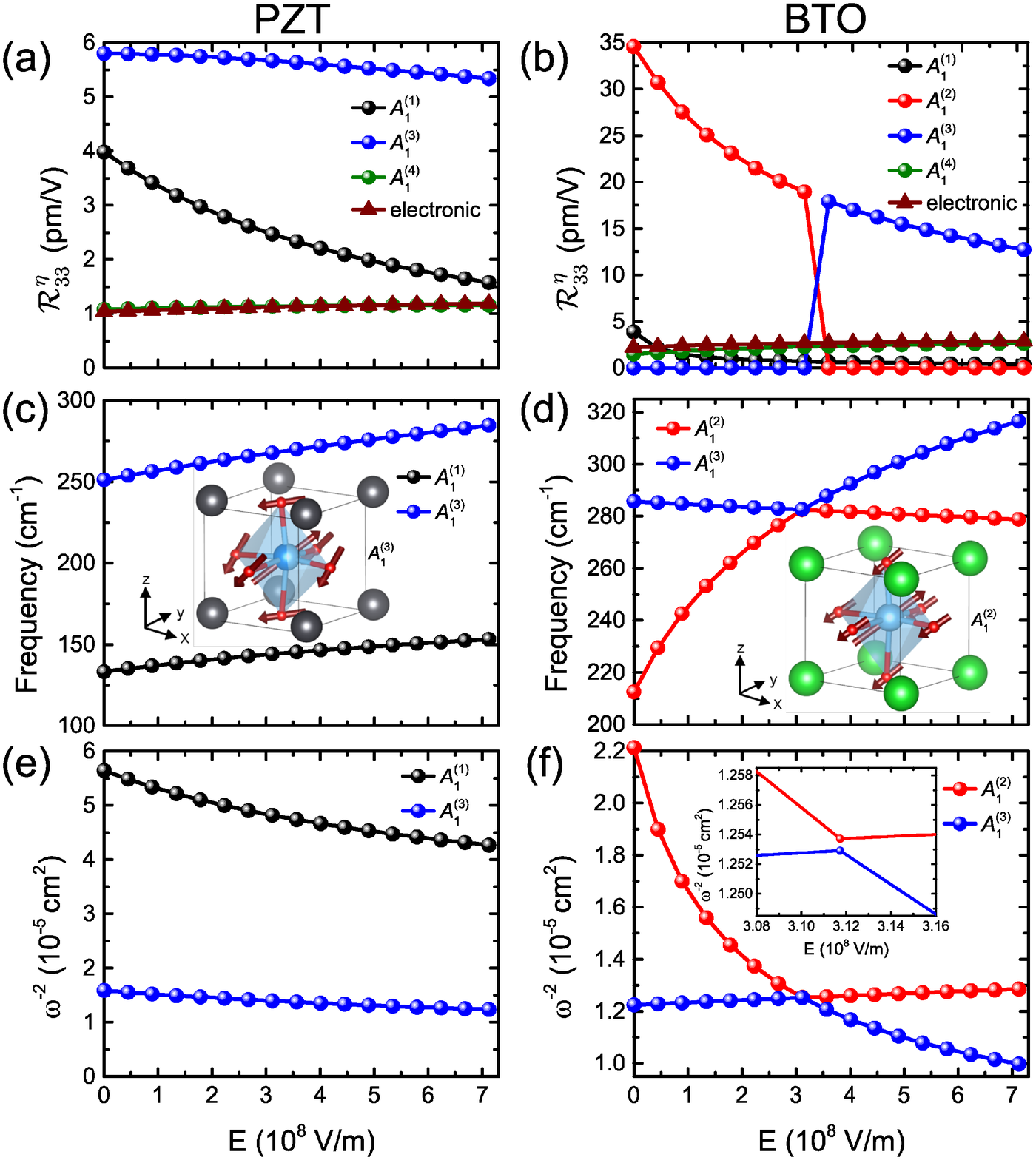}

\caption{Mode decomposition of the clamped EO coefficient ${\mathcal{R}}_{33}^{\mathrm{\eta}}$
in (a) PZT and (b) BTO. Phonon frequency for selected modes at the
$\Gamma$ point of the first Brillouin zone with the insets corresponding
to the eigenvector of $A_{1}^{(3)}$ and $A_{1}^{(2)}$ modes at zero
field in (c) PZT and (d) BTO, respectively. The inverse of the square
of the phonon frequency, $\omega^{-2}$, as a function of electric
field in (e) PZT and (f) BTO. The inset of panel (f) zooms in the
data for electric field between 3.08 $\times$ 10$^{8}$ V/m and 3.16
$\times$ 10$^{8}$ V/m.\label{fig:decomposed contribution and phonon frequency}}
\end{figure}

In order to understand the origin of the linear electro-optic response
in PZT \textit{versus} the nonlinear EO behavior in BTO, we determined
the contribution of each zone-center phonon mode for the ${\mathcal{R}}_{33}^{\eta}$
coefficient, as well as the density functional perturbation theory
(DFPT)-predicted frequency of these modes (see Fig. S3 of the SM \cite{SM}
for all these zone-center phonons), for each of these two systems
and for each investigated electric field. For PZT, Fig.~\ref{fig:decomposed contribution and phonon frequency}(a)
reveals that ${\mathcal{R}}_{33}^{\eta}$ mostly arises from the $A_{1}^{(1)}$
and especially $A_{1}^{(3)}$ modes, with these two modes having frequencies
{[}see Fig.~\ref{fig:decomposed contribution and phonon frequency}(c){]}
behaving in such a manner that the $\omega^{-2}$ inverse of their
square is only weakly (and nearly linearly dependent) on the applied
electric field -- as evidenced in Fig.~\ref{fig:decomposed contribution and phonon frequency}(e).
Incorporating such latter fact when looking again at Eq. (\ref{eq:EO tensor clamped})
naturally explains why ${\mathcal{R}}_{33}^{\eta}$ is mostly independent
of the electric field, that is why the EO response of PZT system is
basically linear (with the slight change of $\omega^{-2}$ with field
generating a weak second-order EO response). Note that the electronic
part of the clamped EO tensor (first term of Eq. (\ref{eq:EO tensor clamped}))
is found to be small, as revealed by Fig.~\ref{fig:decomposed contribution and phonon frequency}(a)
-- therefore indicating the predominant role of ionic contributions
for ${\mathcal{R}}_{33}^{\eta}$ of PZT {[}the same finding holds
for BTO, as demonstrated in Fig.~\ref{fig:decomposed contribution and phonon frequency}(b){]}.
Note also that, as indicated in the inset of Fig.~\ref{fig:decomposed contribution and phonon frequency}(c),
the Ti/Zr ions move along the spontaneous polarization {[}111{]} pseudo-cubic
direction while the three oxygen ions move along the {[}$\bar{1}$$\bar{1}$0{]},
{[}$\bar{1}$0$\bar{1}${]} and {[}0$\bar{1}$$\bar{1}${]} directions,
respectively, in the $A_{1}^{(3)}$ mode of PZT.

In contrast, for BTO, ${\mathcal{R}}_{33}^{\eta}$ takes most of its
value from the $A_{1}^{(2)}$ mode for fields smaller than $\simeq$
3.1 $\times$ 10$^{8}$ V/m, with this mode having a frequency strongly
increasing, and thus an inverse of the square of such frequency strongly
decreasing, with the field. Consequently and according to Eq. (\ref{eq:EO tensor clamped})
also, the electro-optic response of BTO is highly nonlinear, and is
significantly reduced, for fields smaller than $\simeq$ 3.1 $\times$
10$^{8}$ V/m. Note also that the \textit{nonlinear} behavior of $\omega^{-2}$
with field is the culprit behind the existence of \textit{third-order}
EO coefficient. The inset of Fig.~\ref{fig:decomposed contribution and phonon frequency}(d)
shows that Ti ions move along the {[}111{]} direction while the O
ions are displaced along the {[}$\bar{1}$$\bar{1}$$\bar{1}${]}
direction in the $A_{1}^{(2)}$ mode. Furthermore and as evidenced
in Figs.~\ref{fig:decomposed contribution and phonon frequency}(d)
and \ref{fig:decomposed contribution and phonon frequency}(f), a
striking anticrossing between the $A_{1}^{(2)}$ and $A_{1}^{(3)}$
modes then occurs in BTO for a field equal to $\simeq$ 3.1 $\times$
10$^{8}$ V/m. Such anticrossing results in these modes repelling
each other near this critical field, therefore yielding a gap between
these two phonon frequencies and thus between the inverse of their
square, as clearly seen in the inset of Fig.~\ref{fig:decomposed contribution and phonon frequency}(f).
This anticrossing also leads to the eigenvectors of these two modes
inverting their atomic character before \textit{versus} after this
critical field \cite{Jiang2019}. It also makes the $A_{1}^{(3)}$
mode the dominant one for ${\mathcal{R}}_{33}^{\eta}$ above this
critical field, with the resulting EO response being still nonlinear
-- since the frequency of the $A_{1}^{(3)}$ mode also strongly depends
on the magnitude of the field above $\simeq$ 3.1 $\times$ 10$^{8}$
V/m.

One can thus safely conclude that the linear character of the EO response
in PZT \textit{versus} the nonlinear electro-optic response of BTO
mainly originate from the different behavior that the $\omega^{-2}$
of their corresponding predominant modes adopt in these two important
perovskite oxides.

\begin{figure}
\includegraphics[width=8cm]{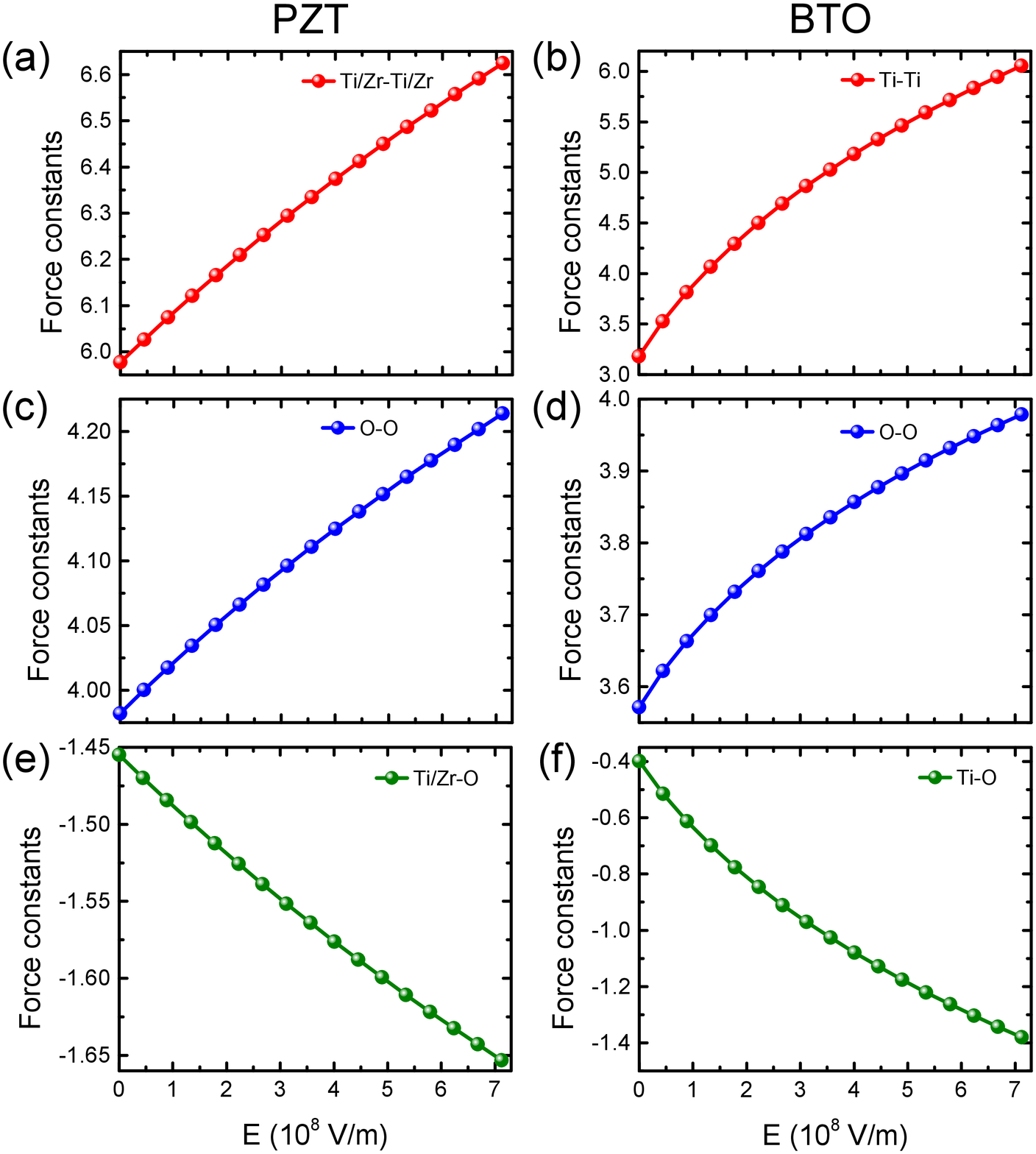}

\caption{Force constants at the $\Gamma$ point of the first Brillouin zone
as a function of electric field for (a) Ti/Zr-Ti/Zr, (c) O-O and (e)
Ti/Zr-O bonds in PZT; and (b) Ti-Ti, (d) O-O, and (f) Ti-O bonds in
BTO.\label{fig:force constant}}
\end{figure}

The next issue to address is therefore to understand why these $\omega^{-2}$
behave in a different manner in PZT and BTO. For that, we reported
the electric-field dependence of the force constants Ti/Zr-Ti/Zr,
O-O and Ti/Zr-O bonds of PZT in the left column of Fig.~\ref{fig:force constant},
and of the force constants of Ti-Ti, O-O and Ti-O of BTO in the the
right column of Fig.~\ref{fig:force constant}. The choice to concentrate
on these specific force constants (rather than those involving Pb
or Ba ions) stems from the atomic character of the eigenvectors associated
with the $A_{1}^{(3)}$ mode in PZT and the $A_{1}^{(2)}$ mode in
BTO {[}see again the inset of Figs.~\ref{fig:decomposed contribution and phonon frequency}(c)
and \ref{fig:decomposed contribution and phonon frequency}(d), respectively{]}.
The selected force constants of Fig.~\ref{fig:force constant} show
nearly linear \textit{versus} strongly nonlinear behaviors as a function
of electric field in PZT and BTO, respectively, which therefore connects
(and explains) the different nature of the macroscopic electro-optic
response in these two systems to some specific atomistic bond characteristics.

In summary, a first-principle technique is developed to tackle nonlinear
electro-optic response of materials at an \textit{ab-initio} level
for the first time, to the best of our knowledge. This method simply
consists of first employing the development of Ref. \cite{Fu2003}
to determine the crystal and atomic structure induced by electric
fields and then use such structure as input of the method of Refs.
\cite{Veithen2004,Veithen2005} to extract EO coefficients as a function
of electric field (note that other atomistic methods, such as those
of Refs. \cite{Umari2002,Sai2002}, can be used to extract the field-induced
structure). This method is presently applied to the $R3m$ phase of
PZT and BTO ferroelectric perovskite oxides, and is also found to
provide similar results than another, more brute-force technique further
proposed and explained in the SM \cite{SM}. Both of these methods
reproduce a recent striking experimental finding, that is why the
EO response of PZT and BTO is linear \textit{versus} nonlinear, respectively
\cite{Chen2014} (note that the SM \cite{SM} also shows that other
optical properties can behave in a different qualitative way between
these two important materials). The scheme indicated in this manuscript
also naturally reveals that it is the field-induced behavior of the
frequency of some specific phonon modes and of some force constants
that are responsible for the difference in nature for the conversion
between electric and optical properties in PZT and BTO. We thus hope
that the present study enhances the knowledge of light-matter interactions
and functional materials, and will also motivate the development of
other techniques allowing the investigation of complex interplay between
light and physical properties. A particular advantage of the proposed
method is that it can be easily employed for the quest of materials
with large nonlinear EO response. 
\begin{acknowledgments}
This work is supported by the National Natural Science Foundation
of China (Grants No.\ 11804138 and No.\ 11825403), Shandong Provincial
Natural Science Foundation (Grant No.\ ZR2019QA008), China Postdoctoral
Science Foundation (Grant No.\ 2018M641905), the Qing Nian Ba Jian
Program, Postdoctoral International Exchange Program of Academic Exchange
Project, Fudan University Super Postdoctoral Program, and Shanghai
Post-doctoral Excellence Program. C.P.\ and L.B.\ thank the ARO
Grant No.\ W911NF-16-1-0227. L.B.\ also acknowledges the DARPA grant
HR0011-15-2-0038 (MATRIX program). 
\end{acknowledgments}

\end{document}